\documentclass[prl,aps,twocolumn,superscriptaddress,showpacs,floatfix]{revtex4}
\usepackage{graphicx}
\usepackage{subfigure}
\usepackage{amsmath, amssymb}
\usepackage{epstopdf}
\usepackage{dcolumn}
\usepackage{bm}

\begin{document}


\title{Macroscopic signature of protected spins in a dense frustrated magnet}

\author{S. Ghosh$^1$}
\email{sghosh@ucmerced.edu}

\author{T. F. Rosenbaum$^2$}

\author{G. Aeppli$^3$}


\affiliation{$^1$School of Natural Sciences, University of California, Merced, CA 95343\\
$^2$James Franck Institute and Department of Physics, University of
Chicago, Chicago, IL 60637\\
$^3$London Centre for Nanotechnology and Department of Physics and
Astronomy, UCL, London, WC1E 6BT,
 UK}


\date{\today}

\begin{abstract}

The inability of systems of interacting objects to satisfy all
constraints simultaneously leads to frustration. A particularly
important consequence of frustration is the ability to access
certain protected parts of a system without disturbing the others.
For magnets such ``protectorates'' have been inferred from theory
and from neutron scattering, but their practical consequences have
been unclear. We show that a magnetic analogue of optical
hole-burning can address these protected spin clusters in a
well-known, geometrically-frustrated Heisenberg system, Gadolinium
Gallium Garnet. Our measurements additionally provide a resolution
of a famous discrepancy between the bulk magnetometry and neutron
diffraction results for this magnetic compound.

\end{abstract}

\maketitle

One important property of frustrated systems is that the number of
most probable (lowest energy) configurations is not finite and grows
exponentially with the system size, resulting in finite
zero-temperature (\textit{T}) entropy. A number of real magnets,
which are among the simplest physical realizations of frustrated
systems exhibit non-zero entropy \cite{bramwell,ramirez1,ramirez2}
as \textit{T}$\rightarrow$0. What has been missing is a simply
measured, macroscopic, dynamical effect associated directly with
decoupled spin subsets. We report here the discovery of such an
effect in a dense, ordered network of spins. Our result shows that
magnetic degrees of freedom can be `protected' \cite{laughlin} from
 each other in a dense magnetic medium.  It also provides a sharp distinction between a well-known magnet, Gadolinium Gallium Garnet,
 where the incompatible constraints are due to lattice geometry,
and conventional spin glasses, where they are due to disorder.\par

\begin{figure}

\includegraphics[scale=0.8]{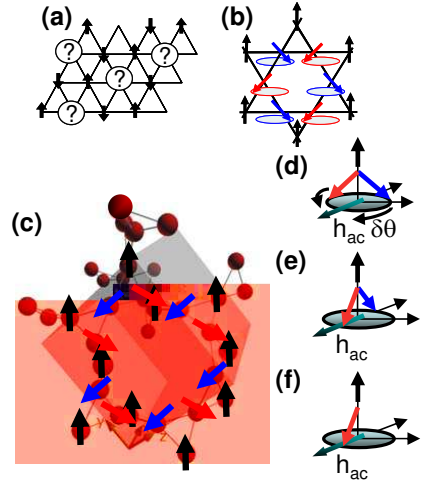}

\caption{(a) Spins at vertices of a triangular Ising
antiferromagnetic (AFM) lattice (labeled ``?'') are geometrically
frustrated  (b) Heisenberg AFM on a Kagom\'{e} lattice. The
``local'' degree of freedom corresponds to a collective rotation of
the red and blue spins around the black spin. (c)  One of the Gd
sublattices of GGG, showing the ten-membered ring of spins which can
rotate about the black spin direction, forming a local degree of
freedom analogous to that in (b). (d-f) Schematics of spin triads to
understand response of local degrees of freedom to small external
fields \textit{h}$_{ac}$. The black spin sublattice is assumed to be
rigid. (a) and (b) are adapted from \cite{aeppli}.}

\label{fig1}

\end{figure}

The archetypical frustrated magnet is the triangular antiferromagnet
(AFM) (Fig. 1a) consisting of Ising spins. If, on a particular
triangle, a spin is antiparallel to its two neighbors, up and down
orientations have equal energy. This spin is a protected degree of
freedom because it can flip as if it were decoupled from its
environment. Such degrees of freedom involving larger groupings of
spins can be found for more complicated lattices (Fig. 1b)
\cite{husimi,obradors,aeppli,lee}. Celebrated three-dimensional
generalizations of the Kagom${\acute{e}}$ lattice include
pyrochlores \cite{taguchi} and garnets. Here we focus on Gadolinium
Gallium Garnet (Gd${_3}$Ga${_5}$O${_12}$ or GGG), a cubic compound
containing magnetic Gd ions on corner-sharing triangles that form
two interpenetrating networks (Fig. 1c). The Weiss temperature
$\Theta{_\textsc{weiss}}$, proportional to a combination \textit{J}
of the exchange (1.5 K) and dipolar ($\sim0.7$ K) couplings between
spins, is approximately 2 K \cite{kinney}.The single ion anisotropy
is less than 0.040 K, making this a Heisenberg system. The geometric
frustration postpones magnetic order to $\textit{T}<$0.180 K, where
the bulk magnetic response \cite{schiffer1} begins to resemble that
of a magnetic glass. However, muon spin relaxation does not display
the behavior conventionally associated with spin glasses below
T${_g}$ \cite{dunsinger}. Additionally, neutron scattering
\cite{pentrenko} reveals sharp magnetic diffraction peaks superposed
on a spin liquid-like structure factor. This, together with recent
theory \cite{yavorskii}, suggests that in the pure limit,
conventional long range AFM order, rather than a disordered spin
glass, coexists with the spin fluid that arises from local degrees
of freedom as illustrated in Fig. 1c.

\begin{figure}

\includegraphics[scale=1.0]{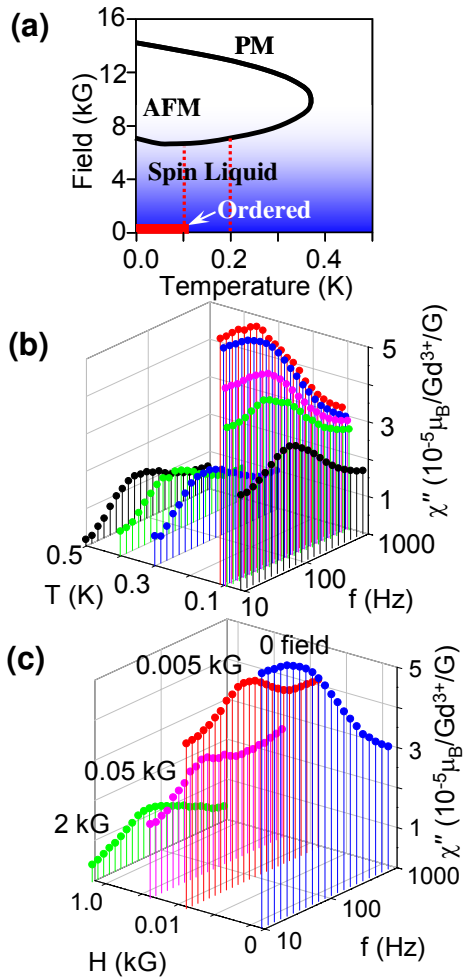}

\caption{(a) Phase diagram (following \cite{schiffer1}). Our study
focuses on the spin liquid phase with \textit{H }$<$4 kG$<6$ kG. The
red dashed lines indicate the temperatures along which we have
conducted detailed \textit{dc} (\textit{H}) studies. (b) $\chi''$,
as a function of
 \textit{T} and \textit{f} at \textit{H} = 0 in
the linear regime. (c) \textit{H} dependence of $\chi''$. The low
\textit{f} plateau in $\chi''(f)$ connected to spin ordering at
0.100 K disappears rapidly with the application of \textit{H}.}

\label{fig2}

\end{figure}

\par
To probe the true
nature of the ground state (Fig. 2a)
 we measured the
dynamic magnetization \textit{M} for oscillating drive fields
${h_{ac}cos(2\pi\textit{ft})}$ down to \textit{T}=0.050 K as a
function of the amplitude ${h_{ac}}$ and frequency \textit{f}. The
sample is a single crystal of GGG (long axis [100]) grown by the
Czochralski method and obtained from Princeton Scientific. In the
limit of small ${h_{ac}}$, \textit{M} is a linear function of
${h_{ac}}$ and the magnetic susceptibility
$\chi(\textit{f})$=\textit{M}/\textit{h}${_ac}$. In fig. 2(b) we
plot the imaginary part $\chi''(\textit{f})$ against \textit{f} and
\textit{T} for ${\textit{h}_{ac}}$=0.04 G, which we have established
(see below) to be in the linear regime. At the high temperature end
(0.500 K, black), $\chi''(\textit{f})$ has a peak at 80 Hz with a
FWHM of $\sim$1.3 decades in \textit{f}, implying that the dynamic
response is described by a distribution of relaxation times
\cite{debye}. As \textit{T} is lowered, the peak height of
$\chi''(\textit{f})$ increases while the basic shape and
\textit{f}$_{peak}$ remain almost the same, until 0.110 K, where the
shape alters drastically. There is no longer a peak in
$\chi''(\textit{f})$, which plateaus at low \textit{f}. The flat,
\textit{f}-independent tail for \textit{f}$\rightarrow$0 originates
from 1/\textit{f} noise in \textit{M}, which is characteristic of
the dynamical scale invariance characteristic of second order phase
transitions as well as certain ordered phases with complex
hierarchies of ground states, such as spin glasses \cite{occio}.
Cooling below 0.100 K reduces $\chi''(\textit{f})$ and its shape
returns to that seen above 0.110 K, again with
\textit{f}$_{peak}$=80 Hz. This is neither expected nor seen for
conventional spin glasses \cite{occio}. What we observe looks more
like a transition to a conventional ordered state, where there is
critical slowing down and condensation of slow modes at the phase
transition, below which the excitations harden as the order grows.
The picture given by the linear susceptibility is consistent with
neutron scattering data \cite{pentrenko}, where relatively sharp
magnetic diffraction peaks correspond to partial AFM order for
\textit{T}$<$\textit{T}$_{N}=0.100$ K$<0.140$ K. However, our bulk
measurements reveal something unusual, that the characteristic
frequency below which a scale-invariant response is observed
corresponds to an energy 0.3 peV$\ll$$k_{B}T_{N}\sim10
\mu$eV$\ll$\textit{J}$\sim$200 $\mu$eV.  For an ideal soft mode
transition, all of these energy scales would be of the same order.
However, soft modes are often accompanied by a `central peak'
\cite{axe}, typically thought to be a contribution of very slowly
relaxing clusters of ordered material in a disordered matrix above
\textit{T}$_{N}$. \par

\begin{figure}

\includegraphics[scale=1.0]{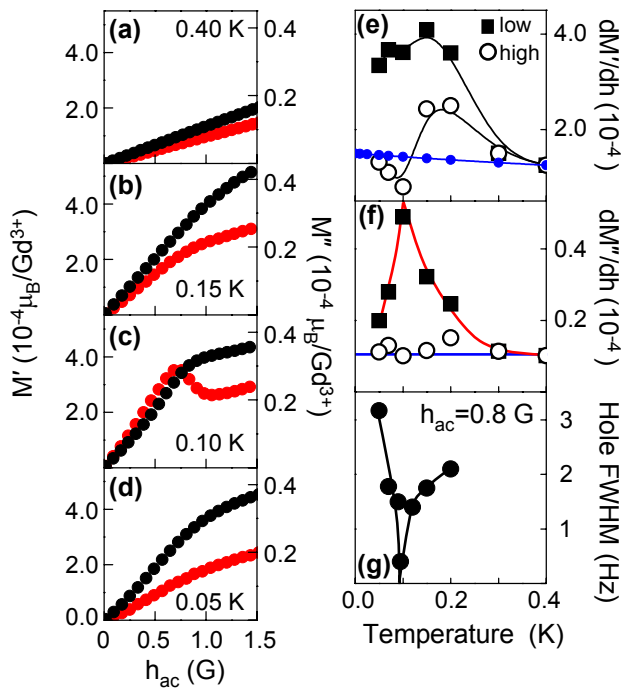}

\caption{(a-d) Real (\textit{M}$'$,black, left axis) and imaginary
(\textit{M}$''$,red, right axis) parts of the magnetization as a
function of \textit{h}$_{ac}$ at different\textit{ T} for
\textit{f}=29.9 Hz. (e) The \textit{T} dependence of
d\textit{M}$'$/d\textit{h}$_{ac}$ at low (\textit{h}$_{ac}$=0.04 G)
and high (\textit{h}$_{ac}$=1.4 G) drive fields. The line (blue,
with solid circles) corresponds to the Curie-Weiss law representing
the extrapolated high-temperature susceptibility dominated by the
frustrated units of GGG.  (f) The analogous plot for \textit{M}$''$.
(g) The spectral width of the hole in the susceptibility as a
function of temperature. All lines are guides to the eye.}

\label{fig3}

\end{figure}

Given the previous work \cite{schiffer1} on the glass-like behavior
of GGG, it is natural to ask whether our data are in the linear
regime. Fig. 3a-d shows measurements of
\textit{M}(\textit{h}$_{ac}$) for \textit{f}=29.9 Hz at a variety of
temperatures. At 0.4 K we are clearly in the linear regime while at
\textit{T}$_{N}$=0.100 K, there are obvious signs of a crossover
from a large, low-\textit{h}$_{ac}$ response to a substantially
smaller high-\textit{h}$_{ac}$ magnetic response. At the same time,
the imaginary part of the magnetization has a strong peak, which
means that dissipation is reduced on going into the
high-\textit{h}$_{ac}$ regime. The highly non-linear response of GGG
revealed at low \textit{h}$_{ac}$ in the current experiment clearly
needs to be taken into account when analyzing magnetic
susceptibility data. Indeed, the AFM phase transition seen in the
neutron experiments and in our linear susceptibility data can be
masked by the non-linearities which clearly matter for ac
magnetometry with \textit{h}$_{ac}$=1 G \cite{schiffer1}.\par
Similar behavior \cite{ghosh} has been seen in the dilute Ising salt
LiHo$_{x}$Y$_{1-x}$F$_{4}$, but with the crucial difference that
there, at high \textit{h}$_{ac}$, \textit{M}$''\rightarrow$0 and
saturation and phase locking are complete.  By contrast, in
concentrated GGG there is a persistent linear (in \textit{h}$_{ac}$)
background. Because we were unable to derive an analytic expression
for \textit{M}(\textit{h}$_{ac}$), we have simply evaluated the low
and high-\textit{h}$_{ac}$ derivatives,
d\textit{M}$'$/d\textit{h}$_{ac}$and
d\textit{M}$''$/d\textit{h}$_{ac}$, as a function of \textit{T},
plotted in Figs. 3e and 3f. The high \textit{h}$_{ac}$ response
shows a strong peak above the ordering temperature, moving towards
the extrapolated (from high \textit{T}) Curie-Weiss susceptibility
\cite{kinney}
 of GGG, while the low \textit{h}$_{ac}$ response has its sharp maximum at \textit{T}$_{N}$, with an amplitude corresponding to
 10$\%$ of the Gd ions rotating freely.

\begin{figure}

\includegraphics[scale=1.0]{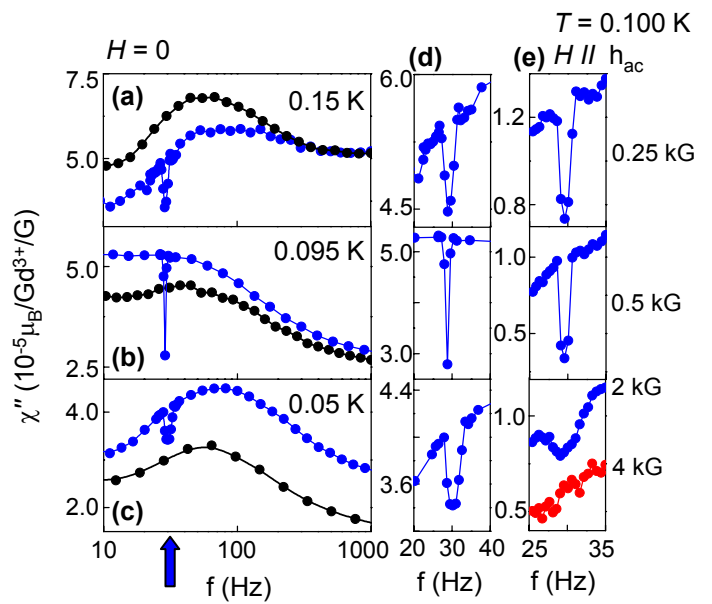}

\caption{(a-c) $\chi''$ as a function of \textit{f} at different
\textit{T } bracketing the ordering temperature in the linear
(black) and non-linear (blue) regimes. The non-linear susceptibility
is measured with a pump-probe combination. \textit{f}$_{pump}$=29.9
Hz (blue arrow). (d) The holes in (a-c) magnified, sharpest at
ordering. (e) Hole-burning at 0.100 K at nonzero \textit{H}. The
ability to burn holes disappears by \textit{H}=4 kG (data shown in
the bottom panel of (e) with an added offset for clarity). }

\label{fig4}
\end{figure}

 When faced with a new dynamical phenomenon in a system with many degrees of freedom, it is important to determine whether it is due to the
 relaxation of many coupled spins, or to the motion of smaller groups of decoupled spins. The
  classic experimental method for discrimination between the two possibilities is hole burning \cite{moerner}. We apply
  simultaneously a sinusoidal pump,
 with amplitude 0.8 G, sufficient to approach the crossover to the high \textit{h}$_{ac}$  regime in Fig. 3a-d, and a small
 amplitude probe (0.04 G).
 Fig. 4a shows how at 0.150 K, the pump carves out a hole in  $\chi''(f )$, centered at \textit{f}$_{pump}$ and with FWHM $\sim$3 Hz.
 No other part of  $\chi''(f )$ is similarly affected although at \textit{f} $<$ 250 Hz the entire spectrum is attenuated relative to that observed
 without pumping. The holes contain about 0.5$\%$
 of the total spectral weight seen at 0.150 K, a macroscopic number $\sim10^{17}$ spins cm$^{-3}$. We can burn similar holes at any frequency \textit{f}
 $<\textit{f}_{peak}$.
 At  0.095 K (Fig. 4(b)), just below \textit{T}$_{N}$, we see a sharper, deeper hole, which then becomes broader and shallower well within
  the ordered phase,
 at 0.050 K. Fig. 3(g) shows how the spectral hole width, inversely proportional to the quality \textit{Q} of the oscillators responsible for the excitations
  at 29.9 Hz, has a sharp minimum at \textit{T}$_{N}$. We have also applied a \textit{dc} bias field \textit{H} to approach the different AFM phase (see Fig. 2c)
  revealed previously above
\textit{ H}=6 kG. As shown in Fig. 4e, the sharp hole is robust up
to at least 0.5 kG, but it is greatly broadened and suppressed at 2
kG and unrecognizable
 at 4 kG.

 \par
 What could give rise to the apparent high-\textit{Q }oscillators in a dense frustrated magnet such as GGG? Most likely are the red and blue spin clusters,
 shown in Fig. 1c, which can be rotated at no cost relative to an ordered, surrounding backbone of black spins.
  A small external field, such as \textit{h}$_{ac}$, will break this rotational symmetry as long as it is not parallel to the field produced by the
   backbone.
   In particular, once the anisotropy field \textit{h}$_{A}$ is exceeded, there will be a spin flop, shown schematically for the spin triad in Fig. 1d,
   with the
   red and blue spins at an angle of 30$^{\circ}$ relative to the black spin and simultaneously orthogonal to \textit{h}$_{ac}$.
   There will then be some small additional
   rotation ($\delta\theta$) of the red and blue spins towards each other and \textit{h}$_{ac}$. In the pure limit,
    where there are no random internal
   fields, there will a much smaller response (Fig. 1e) for \textit{h}$_{ac}<\textit{h}_{A}$ than for \textit{h}$_{ac}>\textit{h}_{A}$. Indeed,
   the imaginary
  part of the bulk response should be rigorously zero \cite{lee2} for \textit{h}$_{ac}\rightarrow0$ for a clean Heisenberg magnet
  (with \textit{h}$_{A}=0$) consisting of AFM triangles.

  Disorder changes the situation markedly and the response can be stronger for \textit{h}$_{ac}<\textit{h}_{A}$ than for \textit{h}$_{ac}>\textit{h}_{A}$.
  For example,
  if there is a missing magnetic atom on the ten-membered ring of Fig. 1c, the ring will have a net, uncanceled moment
  \textit{M}$_{u}$ which is orthogonal to the black spins, and before the spin flop transition occurs the main effect of small \textit{h}$_{ac}$ will
   be to orient  \textit{M}$_{u}$ , a situation which can be understood by simply removing one spin from Fig. 1e to obtain Fig. 1f. Above \textit{h}$_{A}$,
    the spin flop transition will still occur and \textit{M}$_{u}$ will be perpendicular to \textit{h}$_{ac}$, but with a
    correspondingly reduced magnetic response d\textit{M}/d\textit{h}$_{ac}$. This leads to precisely what we observe in Fig. 3. Dissipation is reduced
    at the crossover between the spin-flop regimes because continuous rotation of moments orthogonal
   to \textit{h}$_{ac}$, as illustrated in Fig. 1d, is intrinsically barrier-free, in contrast to the flipping of moments for \textit{h}$_{ac}<\textit{h}_{A}$
   as the applied field oscillates.\par

We can estimate the anisotropy field \textit{h}$_{A}$ to be close to
the value of \textit{h}$_{ac}$ where the susceptibility crosses
between the linear and non-linear regimes and from Fig. 3 this is
approximately 0.8 G. \textit{h}$_{A}$ also leads to an understanding
of the observed characteristic
 frequency \textit{f}$_{peak}$ which  corresponds to the most probable relaxation rate, typically of order J, translating to 50 GHz for GGG.
 Because we are dealing with the relaxation of groups of spins rather than individual moments, the observed \textit{f}$_{peak}$=80 Hz is much smaller.
In this case, the relaxation rate can be estimated using a WKB
approximation \cite{shpyrko} as
\textit{f}$_{peak}$=\textit{f}$_{o}e^{-S/\hbar}$, with the tunneling
 action $\textit{S}/\hbar$=$Ns\sqrt{2k/{\Delta}E}$ for magnetic clusters with \textit{N} spins of magnitude \textit{s}, an anisotropy energy \textit{k}, and an energy barrier
 \textit{$\Delta$E}, due e.g. to interactions with neighboring clusters \cite{shpyrko}. If we set the attempt frequency
 \textit{f}$_{o}$=\textit{J} and \textit{N}=10, the ring size of Fig. 1c, we can obtain \textit{f}$_{peak}$=80 Hz, in agreement with the experiment
 provided that $2k/{\Delta}E\sim1/3$. If we take \textit{k }=0.4 mK =$gs\mu_{B}h_{A}$ , then ${\Delta}E\sim$3 mK,
  which is of the order of the next nearest neighbor exchange interaction \cite{kinney}.

   Inevitable longer-range interactions \cite{yavorskii}, beyond individual triangles, fill out the picture. In particular, at the lowest temperatures the
   fluctuating, uncompensated moments coexist with the unsaturated AFM order revealed by neutrons, and via
   the further neighbor interactions, display dynamics conditioned by the static mean field from the background
   antiferromagnetism.
   If we impose static ferromagnetic correlations via an external field, we see the same reduction in \textit{Q }that is produced by
   lowering \textit{T} below \textit{T}$_{N}$ for \textit{H}=0. At \textit{T}$_{N}$, the static mean field from the other clusters is by definition zero, meaning that the defect-centered
   clusters are truly protected and independent of the AFM backbone, so that here we obtain the highest possible \textit{Q}. On further
    warming, the fluctuation rates of the backbone move through the hole burning frequency, coupling all modes together, and so \textit{Q}
     deteriorates again.\par

     We have produced an experimental resolution of the famous discrepancy between bulk magnetometry
     \cite{schiffer1}
      and magnetic neutron diffraction \cite{pentrenko} for GGG.
      The transition temperature identified previously as the entry point to a spin glass is actually above the condensation
      temperature \textit{T}$_{N}$ for a continuum of ultrasoft modes in the low frequency, linear magnetic response. Completely contrary
      to conventional phase transitions where modes are maximally coupled at a phase transition, at \textit{T}$_{N}$, these oscillators
      achieve maximum decoupling as measured by the quality factors established by magnetic hole burning. Our discovery that the
      characteristic frequency \textit{f}$_{peak}$, for the rollover to a $\chi''(f)$ decreasing with \textit{f} , is independent of \textit{T }and
      so is not thermally activated,
      indicates that the protected, defect-nucleated degrees of freedom are actually behaving as quantum rather than classical objects.
\par
The work at the University of Chicago was supported by DOE BES under
Grant No. DE-FG02-99ER45789. GA acknowledges support from the Wolfson Foundation, EPSRC and
the Rockefeller Foundation for a stay devoted to this manuscript at
the Bellagio Study Center.



\end{document}